# Spin and Orbital Angular Momenta of Electromagnetic Waves in Free Space

Masud Mansuripur

College of Optical Sciences, The University of Arizona, Tucson, Arizona 85721

**Abstract**. We derive exact expressions, in the form of Fourier integrals over the $(\mathbf{k}, \omega)$ domain, for the energy, momentum, and angular momentum of a light pulse propagating in free space. The angular momentum is seen to split naturally into two parts. The spin contribution of each plane-wave constituent of the pulse, representing the difference between its right- and left-circular polarization content, is aligned with the corresponding $k$-vector. In contrast, the orbital angular momentum associated with each plane-wave is orthogonal to its $k$-vector. In general, the orbital angular momentum content of the wavepacket is the sum of an intrinsic part, due, for example, to phase vorticity, and an extrinsic part, $\mathbf{r}_{CM} \times \mathbf{p}$, produced by the linear motion of the center-of-mass $\mathbf{r}_{CM}$ of the light pulse in the direction of its linear momentum $\mathbf{p}$.

**1. Introduction**. It is well known that electromagnetic (EM) waves can carry spin as well as orbital angular momentum [1,2]. The spin angular momentum (SAM) is associated with circular polarization [3-6], while orbital angular momentum (OAM), generally arising in conjunction with the spatial variations of the EM field, is present in optical vortices and vortex-like configurations [7,8]. The problem of separating these two contributions to angular momentum (AM) has been discussed by several authors in the context of both paraxial and non-paraxial beams [1,9-16]. There also exist numerous reports of experimental observations of the two types of AM, as well as methods of generating beams that contain different mixtures of SAM and OAM [1,2,17-25]. For a recent review of the subject including an excellent discussion of the unique interplay between SAM and OAM, the reader is referred to [26].

The goal of the present paper is to demonstrate, using a straightforward yet rigorous Fourier analysis, that the SAM and OAM of an arbitrary light pulse (i.e., wavepacket) are naturally separable in the $(\mathbf{k}, \omega)$ space. Our method, which is firmly rooted in classical electrodynamics, relies solely on the properties of the *E*- and *H*-fields derived from Maxwell's equations. We show that our results are consistent with the well-known decomposition of AM in space-time using the vector potential field, $A(\mathbf{r},t)$, without being prone to criticism due to the gauge-dependence of the vector potential [6,9,11,15].

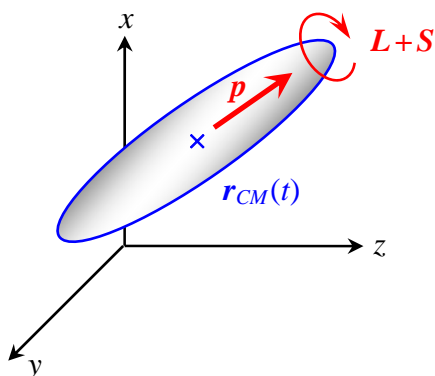

**Fig.1** (color online). A finite-diameter, finite-duration light pulse propagating in free space. The center-of-mass of the wavepacket, $\mathbf{r}_{CM}$, moves with constant velocity in the direction of its linear momentum $\mathbf{p}$. The intrinsic angular momentum of the pulse is the sum of its orbital ($\mathbf{L}$) and spin ($\mathbf{S}$) angular momenta, while the extrinsic part is given by $\mathbf{r}_{CM} \times \mathbf{p}$.

With reference to Fig.1, a finite-diameter, finite-duration light pulse propagating in free space has energy $\mathcal{E}$, momentum $\boldsymbol{p}$, and total angular momentum $\boldsymbol{J}$, all of which are constants of motion. The shape of the pulse changes with time (because of diffraction), but its center-of-mass follows a linear trajectory given by $\boldsymbol{r}_{CM}(t) = \boldsymbol{r}_o + \boldsymbol{v}t$, where $\boldsymbol{r}_o$ is position at time $t = 0$, and the center-of-mass velocity is $\boldsymbol{v} = c^2\boldsymbol{p}/\mathcal{E}$, with $c$ being the speed of light in vacuum. Aside from the trivial (extrinsic) contribution to $\boldsymbol{J}$ of the center-of-mass motion, which, expressed relative to the origin of coordinates, is $\boldsymbol{r}_{CM}(t) \times \boldsymbol{p} = \boldsymbol{r}_o \times \boldsymbol{p}$, the intrinsic AM of the pulse is the sum of its orbital and spin angular momenta, denoted by $\boldsymbol{L} + \boldsymbol{S}$. In the course of the following analysis, these contributions to $\boldsymbol{J}$ will be uniquely and precisely identified in terms of the Fourier representation of the EM field in the $(\boldsymbol{k}, \omega)$ space.

Our method of analysis parallels that of C.G. Darwin in a 1932 paper [4]. The distinction between intrinsic and extrinsic AM, emphasized in recent years [1,10,20], was already apparent in Darwin's analysis, although he did not distinguish the intrinsic orbital momentum from the extrinsic AM associated with the center-of-mass motion. Later authors have either focused their attention on paraxial beams [6,10,13], or tried to investigate the separation of SAM and OAM in space-time domain [9,11,12,14], where the SAM density is expressed as $\varepsilon_o \boldsymbol{E}(\boldsymbol{r},t) \times \boldsymbol{A}(\boldsymbol{r},t)$ – here $\varepsilon_o$ is the permittivity of free space, $\boldsymbol{E}$ is the electric field, $\boldsymbol{A}$ is the vector potential, and $(\boldsymbol{r},t)$ represents space-time coordinates. Our analysis of AM in the Fourier domain shows that restriction to paraxial beams is unnecessary, and that the use of the vector potential, which could be subject to (unfair) criticism due to its gauge dependence, can altogether be avoided. There is also no need to combine classical and quantum mechanical arguments, as has been done, for instance, in [16], in order to isolate or to interpret the various types of AM in the $(\boldsymbol{k}, \omega)$ space.

We will argue in the final section that SAM, being localized in the Fourier domain, should be treated as a global property of the light pulse in space-time. Similarly, OAM is nearly localized in the Fourier domain, in the sense that it depends not only on the local value of the $E$-field in the $(\boldsymbol{k}, \omega)$ domain, but also on the $E$-field gradient at each location. The global nature of OAM in the $(\boldsymbol{r},t)$ space, however, has never been in doubt and need not be emphasized here.

**2. Preliminaries**. Consider a finite-diameter, finite-duration light pulse propagating in free space in an arbitrary direction, as shown in Fig.1. The $E$-field profile of the pulse, being a superposition of homogeneous plane-waves, may be written as follows:

$$\boldsymbol{E}(\boldsymbol{r},t) = \int_{\omega=-\infty}^{\infty} \iint_{k_x^2+k_y^2<(\omega/c)^2} \boldsymbol{\mathcal{E}}(k_x,k_y,\omega)\exp[\mathrm{i}(\boldsymbol{k}\cdot\boldsymbol{r}-\omega t)]\mathrm{d}k_x\mathrm{d}k_y\mathrm{d}\omega. \tag{1a}$$

Here $\boldsymbol{k} = k_x\hat{\boldsymbol{x}} + k_y\hat{\boldsymbol{y}} + k_z\hat{\boldsymbol{z}}$, the real-valued $k$-vector of each plane-wave, satisfies the following constraint imposed by Maxwell's equations for propagation in free space:

$$k_z = (\omega/c)\sqrt{1-(ck_x/\omega)^2-(ck_y/\omega)^2}. \tag{1b}$$

In the above equation, $\omega$ is the temporal frequency of the plane-wave, and $c = 1/\sqrt{\mu_o\varepsilon_o}$ is the speed of light in vacuum, $\mu_o$ and $\varepsilon_o$ being the permeability and permittivity of free space. Since $\boldsymbol{E}(\boldsymbol{r},t)$ is real, its Fourier transform must be Hermitian, that is,

$$\boldsymbol{\mathcal{E}}(-k_x,-k_y,-\omega) = \boldsymbol{\mathcal{E}}^*(k_x,k_y,\omega). \tag{1c}$$

Confinement of the range of $(k_x, k_y)$ to the circle of radius $\omega/c$ in Eq. (1a) is dictated by the need to ensure that the wavepacket is free from evanescent fields. We also deduce from Maxwell's first equation, $\nabla \cdot \boldsymbol{E}(\boldsymbol{r},t) = 0$, that

$$\boldsymbol{k} \cdot \boldsymbol{\mathcal{E}}(k_x, k_y, \omega) = 0. \tag{1d}$$

The $x$- and $y$-components of $\boldsymbol{\mathcal{E}}(k_x, k_y, \omega)$ may be obtained by Fourier transforming the distributions of $E_x(x, y, z = 0, t)$ and $E_y(x, y, z = 0, t)$ as follows:

$$\mathcal{E}_{x,y}(k_x, k_y, \omega) = (2\pi)^{-3} \int_{x=-\infty}^{\infty} \int_{y=-\infty}^{\infty} \int_{t=-\infty}^{\infty} E_{x,y}(x, y, z = 0, t) \exp[-\mathrm{i}(k_x x + k_y y - \omega t)] \mathrm{d}x\mathrm{d}y\mathrm{d}t. \tag{2a}$$

The remaining component $\mathcal{E}_z$ is not independent of $\mathcal{E}_x$ and $\mathcal{E}_y$, and may be obtained from Eq. (1d), namely,

$$\mathcal{E}_z(k_x, k_y, \omega) = -(k_x/k_z)\mathcal{E}_x(k_x, k_y, \omega) - (k_y/k_z)\mathcal{E}_y(k_x, k_y, \omega). \tag{2b}$$

From Eq. (1a), using Maxwell's 3rd equation, $\nabla \times \boldsymbol{E}(\boldsymbol{r},t) = -\partial \boldsymbol{B}(\boldsymbol{r},t)/\partial t$, where $\boldsymbol{B} = \mu_0 \boldsymbol{H}$, we find the magnetic field $\boldsymbol{H}(\boldsymbol{r},t)$ of the light pulse as follows:

$$\boldsymbol{H}(\boldsymbol{r},t) = \int_{\omega=-\infty}^{\infty} \iint_{k_x^2 + k_y^2 < (\omega/c)^2} (\mu_0 \omega)^{-1} \boldsymbol{k} \times \boldsymbol{\mathcal{E}}(k_x, k_y, \omega) \exp[\mathrm{i}(\boldsymbol{k} \cdot \boldsymbol{r} - \omega t)] \, \mathrm{d}k_x \mathrm{d}k_y \mathrm{d}\omega. \tag{3}$$

The above equations may now be used to express the various properties of the light pulse in the form of Fourier integrals; the calculations are tedious but straightforward. We use the following identities to simplify the resulting 9-dimensional integrals:

$$\int_{-\infty}^{\infty} \exp(\mathrm{i}k\zeta)\mathrm{d}\zeta = 2\pi \delta(k), \tag{4a}$$

$$\int_{-\infty}^{\infty} \zeta \exp(\mathrm{i}k\zeta)\mathrm{d}\zeta = -\mathrm{i}2\pi \delta'(k), \tag{4b}$$

$$\int_{-\infty}^{\infty} \delta(k + \utilde{k}) f(\utilde{k}) \mathrm{d}\utilde{k} = f(-k), \tag{5a}$$

$$\int_{-\infty}^{\infty} \delta'(k + \utilde{k}) f(\utilde{k}) \mathrm{d}\utilde{k} = -f'(-k), \tag{5b}$$

$$\boldsymbol{A} \times (\boldsymbol{B} \times \boldsymbol{C}) = (\boldsymbol{A} \cdot \boldsymbol{C})\boldsymbol{B} - (\boldsymbol{A} \cdot \boldsymbol{B})\boldsymbol{C}, \tag{6a}$$

$$(\boldsymbol{A} \times \boldsymbol{B}) \cdot (\boldsymbol{C} \times \boldsymbol{D}) = (\boldsymbol{A} \cdot \boldsymbol{C})(\boldsymbol{B} \cdot \boldsymbol{D}) - (\boldsymbol{A} \cdot \boldsymbol{D})(\boldsymbol{B} \cdot \boldsymbol{C}) \tag{6b}$$

Here $\delta(\cdot)$ is Dirac's delta-function, $\delta'(\cdot)$ is the derivative of $\delta(\cdot)$ with respect to its argument, $f(k)$ is an arbitrary function whose derivative with respect to $k$ is denoted by $f'(k)$, and $\boldsymbol{A}, \boldsymbol{B}, \boldsymbol{C}, \boldsymbol{D}$ are arbitrary real- or complex-valued vectors in 3-dimensional space.

In the course of the following calculations, one encounters the function $\delta(k_z + \utilde{k}_z)$, where $k_z$ is given by Eq. (1b) and $\utilde{k}_z = (\utilde{\omega}/c)\sqrt{1 - (ck_x/\utilde{\omega})^2 - (ck_y/\utilde{\omega})^2}$. (By the time this function is encountered, $\utilde{k}_x$ and $\utilde{k}_y$ in the original expression of $\utilde{k}_z$ have been replaced by $k_x$ and $k_y$, respectively.) The argument of the delta-function, $k_z + \utilde{k}_z$, goes to zero when $\utilde{\omega}$ approaches $-\omega$. Noting that $\partial \utilde{k}_z / \partial \utilde{\omega} \big|_{\utilde{\omega} = -\omega} = \omega/(c^2 k_z)$, and with the aid of Eqs. (4a) and (4b), we find

$$\delta(k_z + \underset{\sim}{k}_z) = (c^2 k_z/\omega)\,\delta(\omega + \underset{\sim}{\omega}), \tag{7a}$$

$$\delta'(k_z + \underset{\sim}{k}_z) = (c^2 k_z/\omega)^2\,\delta'(\omega + \underset{\sim}{\omega}). \tag{7b}$$

These are all the relations needed to arrive at the formulas described in the following sections.

**3. Energy and center of mass**. One way to calculate the total energy $\mathcal{E}$ of the light pulse is to integrate the $z$-component of its Poynting vector $\boldsymbol{S}(\boldsymbol{r},t) = \boldsymbol{E}(\boldsymbol{r},t) \times \boldsymbol{H}(\boldsymbol{r},t)$ over the entire $xy$-plane and over all time. We find

$$\mathcal{E} = \int (\mu_\text{o}\omega')^{-1}\,\boldsymbol{\mathcal{E}}(k_x,k_y,\omega) \times [\boldsymbol{k}' \times \boldsymbol{\mathcal{E}}(k_x',k_y',\omega')]\Big|_z$$
$$\times \exp\{\text{i}[(\boldsymbol{k}+\boldsymbol{k}')\cdot\boldsymbol{r} - (\omega+\omega')t]\}\,\text{d}k_x\text{d}k_y\text{d}\omega\,\text{d}k_x'\text{d}k_y'\text{d}\omega'\text{d}x\text{d}y\text{d}t$$
$$= (2\pi)^3 \varepsilon_\text{o} c \iiint (ck_z/\omega)\boldsymbol{\mathcal{E}}\cdot\boldsymbol{\mathcal{E}}^*\,\text{d}k_x\text{d}k_y\text{d}\omega. \tag{8}$$

Note the obliquity factor $ck_z/\omega$ that multiplies $|\boldsymbol{\mathcal{E}}|^2$ in the final expression for energy. This is just the cosine of the angle between the plane-wave's $k$-vector and the $z$-axis, accounting for the difference between the actual cross-sectional area of the plane-wave and its footprint on the $xy$-plane.

An alternative method of calculating $\mathcal{E}$ is based on integrating the energy densities of the $E$- and $H$-fields over the entire pulse volume, namely,

$$\mathcal{E} = \iiint \left[\tfrac{1}{2}\varepsilon_\text{o}|\boldsymbol{E}(\boldsymbol{r},t)|^2 + \tfrac{1}{2}\mu_\text{o}|\boldsymbol{H}(\boldsymbol{r},t)|^2\right]\text{d}x\text{d}y\text{d}z. \tag{9}$$

Carrying out the integrals in Eq.(9) yields precisely the same result as in Eq.(8), confirming the equivalence of the two methods. The second approach, however, yields the additional result that the total $E$-field energy of the pulse is equal to its $H$-field energy. Also evident in this approach is the fact of energy conservation, as the final expression of $\mathcal{E}$ turns out to be time independent.

The center-of-mass (or center-of-energy) of the wavepacket, $\boldsymbol{r}_{CM}(t)$, is obtained by multiplying the integrand of Eq.(9) with the local position vector $\boldsymbol{r}$, then normalizing the result of integration by the total energy $\mathcal{E}$ of the pulse. We obtain

$$\mathcal{E}\boldsymbol{r}_{CM}(t) = \text{i}(2\pi)^3 \varepsilon_\text{o} c^2 \iiint \{(c/\omega)^2 k_z(\boldsymbol{\mathcal{E}}\cdot\partial_\omega\boldsymbol{\mathcal{E}}^*)\boldsymbol{k} + (k_z/\omega)[(\boldsymbol{\mathcal{E}}\cdot\partial_{k_x}\boldsymbol{\mathcal{E}}^*)\hat{\boldsymbol{x}} + (\boldsymbol{\mathcal{E}}\cdot\partial_{k_y}\boldsymbol{\mathcal{E}}^*)\hat{\boldsymbol{y}}]\}\,\text{d}k_x\text{d}k_y\text{d}\omega$$
$$+ (2\pi)^3 \varepsilon_\text{o} c^2 \left[\iiint (c/\omega)^2 k_z(\boldsymbol{\mathcal{E}}\cdot\boldsymbol{\mathcal{E}}^*)\boldsymbol{k}\,\text{d}k_x\text{d}k_y\text{d}\omega\right]t. \tag{10}$$

The first term on the right-hand-side of Eq.(10) gives the location of the center-of-mass at $t = 0$; note the appearance in the integrand of the partial derivatives of $\boldsymbol{\mathcal{E}}(k_x,k_y,\omega)$ with respect to its three arguments. The second term is a linear function of time, describing the motion of the center-of-mass of the pulse as it propagates through space – while changing shape due to diffraction. The coefficient of time in the second term of Eq.(10), when divided by $c^2$ (to convert energy to mass) should be the linear momentum of the wavepacket – this will be verified in Sec.4 where we calculate the momentum directly. We should also mention that, during calculations that led to Eq.(10), we found the center-of-mass of the $E$-field to coincide with that of the $H$-field at all times $t$.

The following simple exercise should clarify the roles played by each of the three terms in the first integral of Eq.(10), as well as their respective contributions to the value of $r_{CM}(0)$. Suppose the light pulse is shifted by $(x_o, y_o)$ in the $xy$-plane, and also delayed by $t_o$ along the time axis. The Fourier transform of $E_{x,y}(x - x_o, y - y_o, z = 0, t - t_o)$ will then be given, in accordance with Eq.(2a), by $\exp[-\mathrm{i}(k_x x_o + k_y y_o - \omega t_o)]\mathcal{E}_{x,y}(k_x, k_y, \omega)$. As a result of this shift, $\mathcal{E} \cdot \partial_\omega \mathcal{E}^*$ will change to $\mathcal{E} \cdot (\mathrm{i} t_o \mathcal{E}^* + \partial_\omega \mathcal{E}^*)$, while $\mathcal{E} \cdot \partial_{k_x} \mathcal{E}^*$ will become $\mathcal{E} \cdot (-\mathrm{i} x_o \mathcal{E}^* + \partial_{k_x} \mathcal{E}^*)$, and $\mathcal{E} \cdot \partial_{k_y} \mathcal{E}^*$ will become $\mathcal{E} \cdot (-\mathrm{i} y_o \mathcal{E}^* + \partial_{k_y} \mathcal{E}^*)$. Substitution into Eq.(10) reveals that the new center-of-mass is shifted by $(x_o, y_o)$ in the $xy$-plane, and also pulled back in proportion to $t_o$ along the propagation path. The remaining term in Eq.(10), however, will not be affected by the shift at all, nor will there be any change in the pulse energy $\mathcal{E}$ of Eq.(8) as a result of this shift.

**4. Momentum of the light pulse**. The linear momentum density of EM fields is given by $S(r,t)/c^2$, where $S$ is the Poynting vector. Integrating this momentum density over the volume of the light pulse yields

$$p = (1/c^2) \int \mathcal{E}(k_x, k_y, \omega) \times [(\mu_o \omega')^{-1} k' \times \mathcal{E}(k'_x, k'_y, \omega')]$$

$$\times \exp\{\mathrm{i}[(k + k') \cdot r - (\omega + \omega') t]\} \, \mathrm{d}k_x \mathrm{d}k_y \mathrm{d}\omega \, \mathrm{d}k'_x \mathrm{d}k'_y \mathrm{d}\omega' \mathrm{d}x \mathrm{d}y \mathrm{d}z$$

$$= (2\pi)^3 \varepsilon_o \int (c/\omega)^2 k_z (\mathcal{E} \cdot \mathcal{E}^*) k \, \mathrm{d}k_x \mathrm{d}k_y \mathrm{d}\omega. \qquad (11)$$

Clearly, the total momentum $p$ is time-independent, confirming momentum conservation as the pulse propagates in space. One of the two $c/\omega$ factors in the final integrand normalizes the $k$-vector, so that the momentum $\varepsilon_o |\mathcal{E}|^2$ associated with each plane-wave is directed along the unit-vector $\hat{\kappa} = c k/\omega$. The other $c/\omega$ normalizes $k_z$ to yield the obliquity factor $c k_z / \omega$, which, as mentioned earlier, accounts for the difference between the footprint of a beam on the $xy$-plane and its cross-sectional area perpendicular to its propagation direction.

Comparing Eq.(11) with Eq.(8), we see that the momentum associated with each plane-wave is equal to the corresponding energy divided by $c$. However, the magnitude $p$ of the total momentum is generally less then $\mathcal{E}/c$, because of the spread in the direction of the $k$-vectors. In other words, the group velocity (which is the velocity of the center-of-mass of the pulse) is generally less than $c$, approaching $c$ only when the beam becomes highly paraxial.

**5. Angular momentum**. The angular momentum density of the EM field with respect to a reference point $r_{\mathrm{ref}}$ is given by $(r - r_{\mathrm{ref}}) \times S(r,t)/c^2$. The total AM of the pulse with respect to the origin ($r_{\mathrm{ref}} = 0$) is thus found to be

$$J = (1/c^2) \int r \times \{\mathcal{E}(k_x, k_y, \omega) \times [(\mu_o \omega')^{-1} k' \times \mathcal{E}(k'_x, k'_y, \omega')]\}$$

$$\times \exp\{\mathrm{i}[(k_x + k'_x) x + (k_y + k'_y) y + (k_z + k'_z) z - (\omega + \omega') t]\} \, \mathrm{d}k_x \mathrm{d}k_y \mathrm{d}\omega \, \mathrm{d}k'_x \mathrm{d}k'_y \mathrm{d}\omega' \mathrm{d}x \mathrm{d}y \mathrm{d}z$$

$$= \mathrm{i}(2\pi)^3 \varepsilon_o \iiint (c/\omega)^2 \{(\mathcal{E}_x \mathcal{E}_y^* - \mathcal{E}_y \mathcal{E}_x^*) k + k_z [(\mathcal{E} \cdot \partial_{k_x} \mathcal{E}^*) \hat{x} + (\mathcal{E} \cdot \partial_{k_y} \mathcal{E}^*) \hat{y}] \times k \} \mathrm{d}k_x \mathrm{d}k_y \mathrm{d}\omega. \quad (12)$$

Conservation of AM is readily verified by the fact that $J$ is time-independent. The expression of $J$ as an integral over the $(k, \omega)$ domain clearly demonstrates the existence of two contributions to the total AM. The component aligned with the $k$-vector gives rise to SAM, with $(\varepsilon_\text{o} c/\omega)\text{Im}(2\mathcal{E}_x\mathcal{E}_y^*)\boldsymbol{k}/k_z$ representing the difference between right- and left-circular polarization contributions to each plane-wave's angular momentum; see Sec. 6 for more details on this point. The second half of Eq. (12), whose integrand is orthogonal to $\boldsymbol{k}$, represents all other contributions to angular momentum including OAM and $\boldsymbol{r}_{CM}\times\boldsymbol{p}$, which is associated with the center-of-mass motion.

Comparing the second half of Eq. (12) with the expression of $\boldsymbol{r}_{CM}$ in Eq. (10), we observe that the term $(\boldsymbol{\mathcal{E}}\cdot\partial_\omega\boldsymbol{\mathcal{E}}^*)\boldsymbol{k}$ in the integrand of $\boldsymbol{r}_{CM}$ makes no contribution to the angular momentum $\boldsymbol{J}$, presumably because it is aligned with $\boldsymbol{k}$. The remaining term, $(\boldsymbol{\mathcal{E}}\cdot\partial_{k_x}\boldsymbol{\mathcal{E}}^*)\hat{\boldsymbol{x}}+(\boldsymbol{\mathcal{E}}\cdot\partial_{k_y}\boldsymbol{\mathcal{E}}^*)\hat{\boldsymbol{y}}$, however, after being cross-multiplied into the local $k$-vector, fully participates in the expression of $\boldsymbol{J}$ in Eq. (12). We emphasize that, while the first half of the expression of $\boldsymbol{J}$ in Eq. (12) represents the purely intrinsic SAM of the light pulse, its second half contains both intrinsic and extrinsic contributions to the angular momentum [20]. Once the extrinsic part, $\boldsymbol{r}_{CM}\times\boldsymbol{p}$, has been subtracted from the second half of Eq. (12), the remainder will correspond to the OAM of the wavepacket with respect to the center-of-mass, namely, the intrinsic OAM.

Note that there is no a priori reason for $\boldsymbol{J}$ or either of its constituents, $\boldsymbol{L}$ or $\boldsymbol{S}$, to be aligned with the general direction of propagation of the wavepacket, which direction is specified by the linear momentum $\boldsymbol{p}$. Depending on the specific distribution of $\boldsymbol{\mathcal{E}}(k_x,k_y,\omega)$ in the $(\boldsymbol{k},\omega)$ space, $\boldsymbol{J}$, $\boldsymbol{L}$, and $\boldsymbol{S}$ could have very different orientations relative to each other and also relative to $\boldsymbol{p}$.

**6. Degree of circular polarization of a plane-wave.** With reference to Fig. 2, consider a plane-wave propagating along the unit-vector $\hat{\boldsymbol{\kappa}}=c\boldsymbol{k}/\omega=\kappa_x\hat{\boldsymbol{x}}+\kappa_y\hat{\boldsymbol{y}}+\kappa_z\hat{\boldsymbol{z}}$. The polarization state of this plane-wave is a superposition of right- and left-circular components of magnitude $A_R=|A_R|\exp(\text{i}\phi_R)$ and $A_L=|A_L|\exp(\text{i}\phi_L)$, respectively. The degree of circular polarization of the plane-wave, which is intimately related to its SAM, may be defined as $|A_R|^2-|A_L|^2$.

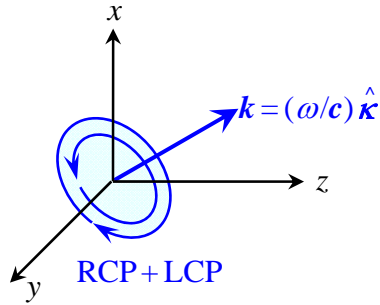

**Fig. 2** (color online). Plane-wave propagating in free space along its $k$-vector, having a mixture of right- and left-circular polarization with amplitudes $A_R$ and $A_L$, respectively.

In order to relate the degree of circular polarization to the $E$-field components $E_x$ and $E_y$, we define a special pair of orthogonal unit-vectors $\hat{\boldsymbol{u}}$ and $\hat{\boldsymbol{v}}$ in the plane of polarization. $\hat{\boldsymbol{u}}$ is confined to the $xy$-plane, that is, $\hat{\boldsymbol{u}}=u_x\hat{\boldsymbol{x}}+u_y\hat{\boldsymbol{y}}$, its orthogonality to $\hat{\boldsymbol{\kappa}}$ yielding

$$\hat{\boldsymbol{\kappa}}\cdot\hat{\boldsymbol{u}}=\kappa_x u_x+\kappa_y u_y=0. \qquad (13)$$

The remaining vector must then be

$$\hat{\boldsymbol{v}} = \hat{\boldsymbol{\kappa}} \times \hat{\boldsymbol{u}} = -\kappa_z u_y \hat{\boldsymbol{x}} + \kappa_z u_x \hat{\boldsymbol{y}} + (\kappa_x u_y - \kappa_y u_x)\hat{\boldsymbol{z}}. \tag{14}$$

The total *E*-field is thus given by

$$\boldsymbol{E} = E_x \hat{\boldsymbol{x}} + E_y \hat{\boldsymbol{y}} + E_z \hat{\boldsymbol{z}} = A_R(\hat{\boldsymbol{u}} - \mathrm{i}\hat{\boldsymbol{v}}) + A_L(\hat{\boldsymbol{u}} + \mathrm{i}\hat{\boldsymbol{v}}). \tag{15}$$

If we now compute $E_x E_y^*$, we will find

$$\begin{aligned} E_x E_y^* &= (1-\kappa_z^2)(|A_R|^2 + |A_L|^2) u_x u_y + 2(1+\kappa_z^2)|A_R||A_L|\cos(\phi_R - \phi_L) u_x u_y \\ &\quad + 2\kappa_z |A_R||A_L|\sin(\phi_R - \phi_L)(u_x^2 - u_y^2) + \mathrm{i}\kappa_z(|A_R|^2 - |A_L|^2). \end{aligned} \tag{16}$$

Clearly, the degree of circular polarization of the plane-wave is given by

$$|A_R|^2 - |A_L|^2 = (\omega/ck_z)\,\mathrm{Im}(E_x E_y^*). \tag{17}$$

Returning now to the expression of $\boldsymbol{J}$ in Eq.(12), we write the SAM part of the integrand as follows:

$$\mathrm{i}\varepsilon_\mathrm{o}(c/\omega)^2(\boldsymbol{\mathcal{E}}_x\boldsymbol{\mathcal{E}}_y^* - \boldsymbol{\mathcal{E}}_y\boldsymbol{\mathcal{E}}_x^*)\boldsymbol{k} = -2(ck_z/\omega)(\varepsilon_\mathrm{o}c/\omega)(\omega/ck_z)\,\mathrm{Im}(\boldsymbol{\mathcal{E}}_x\boldsymbol{\mathcal{E}}_y^*)\hat{\boldsymbol{\kappa}}. \tag{18}$$

Here $ck_z/\omega$ is the obliquity factor mentioned earlier, and $\varepsilon_\mathrm{o}c/\omega$ is the factor that converts the degree of circular polarization, $(\omega/ck_z)\,\mathrm{Im}(\boldsymbol{\mathcal{E}}_x\boldsymbol{\mathcal{E}}_y^*)$, to angular momentum. The factor of 2 can be accounted for by recognizing that each plane-wave's *E*-field amplitude, as defined by Eq.(2a), is equally split between $+\omega$ and $-\omega$. We have thus demonstrated the correspondence between the degree of circular polarization of individual plane-waves and the total SAM as expressed by the first term on the right-hand-side of Eq.(12).

**7. Spin angular momentum and the vector potential**. The vector potential $\boldsymbol{A}(\boldsymbol{r},t)$ is defined as the vector field whose curl is the *B*-field, that is, $\nabla \times \boldsymbol{A}(\boldsymbol{r},t) = \boldsymbol{B}(\boldsymbol{r},t)$. This definition fixes the transverse component of the vector potential, but leaves its longitudinal component unspecified. In the Coulomb gauge, the longitudinal component of $\boldsymbol{A}(\boldsymbol{r},t)$ is set to zero, i.e., $\nabla \cdot \boldsymbol{A}(\boldsymbol{r},t) = 0$. Thus, in the Fourier domain, $\boldsymbol{\mathcal{A}}(\boldsymbol{k},\omega) = \mathrm{i}(c/\omega)^2 \boldsymbol{k} \times \boldsymbol{\mathcal{B}}(\boldsymbol{k},\omega)$. Considering that $\boldsymbol{\mathcal{B}} = \omega^{-1}\boldsymbol{k}\times\boldsymbol{\mathcal{E}}$, the relation between the vector potential and the *E*-field is $\boldsymbol{\mathcal{A}}(\boldsymbol{k},\omega) = -\mathrm{i}\omega^{-1}\boldsymbol{\mathcal{E}}(\boldsymbol{k},\omega)$.

In the literature, the "spin angular momentum density" of EM fields in vacuum is often expressed as $\varepsilon_\mathrm{o}\boldsymbol{E}(\boldsymbol{r},t)\times\boldsymbol{A}(\boldsymbol{r},t)$ [1,6,15]. The total SAM of a light pulse is thus given by

$$\begin{aligned} \boldsymbol{S} &= \iiint \varepsilon_\mathrm{o}\boldsymbol{E}(\boldsymbol{r},t)\times\boldsymbol{A}(\boldsymbol{r},t)\,\mathrm{d}x\mathrm{d}y\mathrm{d}z = -\mathrm{i}\varepsilon_\mathrm{o}\int \boldsymbol{\mathcal{E}}(k_x,k_y,\omega)\times[\omega'^{-1}\boldsymbol{\mathcal{E}}(k_x',k_y',\omega')] \\ &\quad \times \exp\{\mathrm{i}[(k_x+k_x')x + (k_y+k_y')y + (k_z+k_z')z - (\omega+\omega')t]\}\,\mathrm{d}k_x\mathrm{d}k_y\mathrm{d}\omega\,\mathrm{d}k_x'\mathrm{d}k_y'\mathrm{d}\omega'\mathrm{d}x\mathrm{d}y\mathrm{d}z \\ &= \mathrm{i}(2\pi)^3\varepsilon_\mathrm{o}\iiint (c/\omega)^2(\boldsymbol{\mathcal{E}}_x\boldsymbol{\mathcal{E}}_y^* - \boldsymbol{\mathcal{E}}_y\boldsymbol{\mathcal{E}}_x^*)\boldsymbol{k}\,\mathrm{d}k_x\mathrm{d}k_y\mathrm{d}\omega. \end{aligned} \tag{18}$$

The above result is seen to be identical with the first term in the expression of $\boldsymbol{J}$ in Eq.(12), thus confirming the equivalence of the space-time method and the Fourier method of separating SAM from OAM. Separation by means of the vector potential has, on occasion, been criticized on the grounds that it renders the result "gauge dependent." The criticism is unfair, considering that only the transverse component of the vector potential, which *is* gauge invariant, appears in the expressions for SAM and OAM in these analyses. In any case, the alternative derivation of the final result of Eq.(18) via Eq.(12) relies solely on the gauge-invariant *E*- and *H*-fields, and should therefore be immune to such criticism.

**8. Concluding remarks**. Much has been said in recent years concerning the different manifestations of SAM and OAM in classical optical systems. The significance of such distinctions, however, should not be exaggerated, at least in the context of systems governed by classical (as opposed to quantum) electrodynamics. Also, the simplicity of converting SAM to OAM and vice-versa upon reflection from a hollow metallic cone [27], for example, calls for a more nuanced view of the nature of these two types of angular momentum and their intertwined relationship. Considering that both SAM and OAM are non-local in the (*r*,*t*) space, one should exercise caution in relating the results of local measurements to these global properties of the light beam.

As has already been pointed out, the total AM of a light pulse with respect to an arbitrary point $r_\text{ref}$ is the integral of $(r - r_\text{ref}) \times S(r,t)/c^2$ over the spatial volume occupied by the pulse. This prescription applies whether the AM is due to the polarization state of the wavepacket, or its saptial variation (e.g., vorticity), or a mixture of the two. In other words, one does not distinguish SAM from OAM when computing the total angular momentum of an EM wave. Since the distinction cannot be based on an analysis of the Poynting vector profile, it must lie in the local or global properties of the *E*- and *H*-fields, and, perhaps more importantly, in the *methods* of monitoring such properties.

Suppose, for instance, that the field at and around a given point *r* is circularly polarized. If we place at *r* a small spherical particle of an absorptive material, the particle acquires some mechanical AM from the EM field and begins to rotate on its axis. The essential physics of this process involves the appearance (within the particle) of an induced dipole moment *p*, which co-rotates with the local *E*-field. The strength *p* of the dipole-moment is proportional to the local *E*-field, with the proportionality constant being the magnitude of the particle's electric susceptibility. The absorptive nature of the particle renders its susceptibility complex-valued. Absorption thus produces a lag between the induced dipole moment *p* and the local *E*-field, with the phase of the complex susceptibility determining the angle by which the rotating vector *p* lags behind the co-rotating *E*. The torque experienced by the particle is then given by *p*×*E*, which is responsible for the spinning of the particle on its axis. An isotropic and transparent particle would *not* have behaved in this way, because its induced dipole moment *p* would have been aligned with the *E*-field at all times. In contrast, a transparent birefringent particle *would* have picked up some spin from the local field, as its birefringence would produce the all-important angle between the induced dipole *p* and the local *E*-field. In all these examples, the local or global structure of the field's Poynting vector is irrelevant; what matters is that the EM field at point *r* has a net circular polarization, and that the experiment is designed to sense this local polarization state.

When a small absorptive particle takes in energy from the local EM field and converts it to heat, we do not claim that the EM energy has been of a "thermal nature." Nor do we refer to the energy as "radiative" when it is picked up by a transparent particle and radiated back into space. Similarly, the EM energy needed to set small particles in linear or rotational motion is never considered to be of "mechanical" type. The EM energy, of course, is the same in all these cases, but the particular choice of the sensor has revealed its ability to transform itself into different forms. By the same token, one must be careful in drawing a distinction between SAM and OAM as revealed by the particular choice of a sensor of angular momentum, especially when the beam or the wavepacket under consideration happens to be non-paraxial.

Considered from a different perspective, even a linearly-polarized light pulse with no angular momentum whatsoever has the ability to transfer SAM to a small (birefringent) quarter-wave-plate (QWP), provided that the polarization direction of the incident pulse does *not* coincide with the plate's axes. The pulse, of course, does not have any SAM to begin with, but it becomes circularly (or elliptically) polarized upon transmission through the plate. Conservation of AM then ensures that the QWP rotates in the opposite direction. If, instead of the QWP, a spiral phase-plate is used to impart a helical phase to the pulse, the transmitted light will become an optical vortex, and the phase-plate acquires a spinning motion similar to that of the QWP. In both cases the initial beam is devoid of AM, yet the sensor (QWP in the first case, spiral phase-plate in the second) manages to pick up a spinning motion. In no way does the observed behavior of the "sensor" in these examples reflect upon the AM content of the incident wavepacket.

As another example, consider a linearly-polarized beam of light endowed with some degree of vorticity, say, a Laguerre-Gaussian beam with a topological charge of 1; such a beam is said to have some orbital but no spin angular momentum. Let the beam propagate along its axis of symmetry, the *z*-axis, and assume that a small, spherical, absorptive particle is placed in the beam's path at an off-axis location, say, ***r***. (For the sake of simplicity, assume that the particle is somehow trapped in a transverse plane such that, while it can move within the *xy*-plane, its motion along *z* is constrained.) Upon absorbing a fraction of the incident light, the particle will start on a circular path around the *z*-axis, but it will *not* spin on its own axis. Changing the state of polarization of the vortex beam from linear to circular will cause the particle to spin while traveling on its circular orbit around *z*. Reversing the sense of circular polarization will reverse the spin angular momentum of the particle, but will not affect its orbital motion.

Such observations are typically invoked to distinguish electromagnetic SAM from OAM. Of course, what is implied to be distinct about the two types of optical AM in the above situation is the local nature of SAM versus the global nature of OAM. That distinction will no longer be evident if, instead of a small absorptive particle (i.e., local sensor), one placed a disk or a circular ring of an absorptive material (i.e., global sensor) in the light's path. Such a disk or ring, placed perpendicular to and centered on the *z*-axis, would indiscriminately absorb SAM as well as OAM from the incident beam, both of which would set it in rotational motion around *z*.

If, instead of an absorptive disk, one placed a QWP in the path of the above beam (again perpendicular to *z*), those plane-wave constituents of the beam whose *k*-vector deviates from the *z*-axis would not transfer the full amount of their SAM to the wave plate. The method, therefore, is not suitable for measuring the SAM content of a non-paraxial beam. However, allowing the beam to propagate into the far field – where its individual plane-waves spatially separate – then measuring and adding up the SAM content of all its plane-waves, would be an acceptable way of isolating the SAM content of the beam [12]. These considerations highlight the pitfalls of inferring the presence of SAM or OAM from the results of local or even global measurements.

The above arguments also apply to situations where the intensity of the light pulse is continually reduced until each pulse contains a single photon. The behavior of the aforementioned absorptive particle, the birefringent particle, and the absorptive or birefringent ring and disk in response to individual photons that carry SAM and/or OAM is an interesting topic in its own right, but one that is beyond the scope of the present (classical) discussion.

**Acknowledgement**. The author is grateful to Chunfang Li, Ewan Wright, Poul Jessen, and Armis Zakharian for helpful discussions.